\input amstex
\magnification 1200
\TagsOnRight
\def\qed{\ifhmode\unskip\nobreak\fi\ifmmode\ifinner\else
 \hskip5pt\fi\fi\hbox{\hskip5pt\vrule width4pt
 height6pt depth1.5pt\hskip1pt}}
 \def\adots{\mathinner{\mkern2mu\raise1pt\hbox{.}
\mkern3mu\raise4pt\hbox{.}\mkern1mu\raise7pt\hbox{.}}}
\def\sdots{\mathinner{
     \mskip.01mu\raise1pt\vbox{\kern1pt\hbox{.}}
     \mskip.01mu\raise3pt\hbox{.}
     \mskip.01mu\raise5pt\hbox{.}
\mskip1mu}}
\NoBlackBoxes
\baselineskip 20 pt
\parskip 5 pt

\centerline {\bf SYMMETRIES FOR EXACT SOLUTIONS TO}
\centerline {\bf THE NONLINEAR SCHR\"ODINGER EQUATION}

\vskip 10 pt
\centerline {Tuncay Aktosun and Theresa Busse}
\vskip -6 pt
\centerline {Department of Mathematics}
\vskip -6 pt
\centerline {University of Texas at Arlington}
\vskip -6 pt
\centerline {Arlington, TX 76019-0408, USA}

\centerline{Francesco Demontis and Cornelis van der Mee}
\vskip -6 pt
\centerline{Dipartimento di Matematica e Informatica}
\vskip -6 pt
\centerline{Universit\`a di Cagliari}
\vskip -6 pt
\centerline{Viale Merello 92, 09123 Cagliari, Italy}

\vskip 10 pt

\noindent {\bf Abstract}: A certain symmetry is exploited in
expressing exact solutions to the focusing nonlinear
Schr\"odinger equation in terms of a triplet of constant matrices.
Consequently, for any number of bound states with any number
of multiplicities the corresponding soliton solutions are explicitly written
in a compact form in terms of a matrix triplet.
Conversely, from such a soliton solution the corresponding
transmission coefficients, bound-state poles, bound-state norming constants
and Jost solutions for the associated Zakharov-Shabat system
are evaluated explicitly.
It is also shown that these results hold for the matrix
nonlinear
Schr\"odinger equation of any matrix size.


\vskip -6 pt
\par\noindent {\bf Short title:} Symmetries for exact solutions to the NLS equation
\newpage

\noindent {\bf 1. INTRODUCTION}
\vskip 3 pt

Consider the focusing cubic nonlinear Schr\"odinger (NLS) equation
$$iu_t+u_{xx}+2|u|^2u=0,\tag 1.1$$
where subscripts denote the appropriate partial derivatives. It arises in applications
as diverse as wave propagation in nonlinear media [15], surface waves on sufficiently
deep waters [14,15], and signal propagation in optical fibers [10-12]. Its initial-value
problem is known to be solvable by the inverse scattering transform method [1,2,13,15].
In other words, certain solutions to (1.1) can be viewed as a potential
in the Zakharov-Shabat system
$$\displaystyle\frac{d \varphi(\lambda,x,t)}{dx}=\bmatrix -i\lambda&u(x,t)\\
\noalign{\medskip}
-u(x,t)^\ast&i\lambda\endbmatrix \varphi(\lambda,x,t),\tag 1.2$$
where an asterisk is used to denote complex conjugation, and $u(x,t)$ can be recovered
 from $u(x,0)$ with the help of the scattering data sets for
(1.2) at $t=0$ and at time $t.$

Exact solutions to nonlinear partial differential equations are of great
interest. Such solutions may be helpful to better
understand the corresponding nonlinearity, and
they may also be useful in producing testing means to determine
accuracy of numerical methods for solving
nonlinear partial differential equations. This paper is related to
exact solutions to (1.1).

In previous papers [3,6-8] we presented a method to construct exact solutions to
(1.1) that are globally analytic on the entire $xt$-plane and decay exponentially as
$x\to\pm\infty$ at each fixed $t\in{\bold R}$. This has been achieved
by using a matrix
triplet $(A,B,C),$ where all eigenvalues of $A$ have positive
real parts. A similar method was applied to the
Korteweg-de Vries equation on the half-line [4]. The same method
is also applicable to various other nonlinear partial differential
equations that are integrable by the inverse scattering transform
with the help of a Marchenko integral equation.

In this paper we analyze the method of [3] when the eigenvalues of
the matrix $A$ in the triplet $(A,B,C)$ do not all have positive real parts.
It is already know that if one or more eigenvalues of $A$ are purely imaginary, the
corresponding scattering coefficients for (1.2) contain discontinuities at some real
values of $\lambda$ and hence the corresponding $u(x,t)$ cannot be
analytic on the entire $xt$-plane. Furthermore, it is already known that for
soliton solutions to (1.1), the corresponding
transmission coefficients for (1.2) must have a pole and a zero appearing
as a pair located symmetrically with respect to the real axis, which implies
that a pair (or more pairs) of
eigenvalues of $A$ cannot be located
symmetrically with respect to
the imaginary axis.
Thus, in our paper we mainly concentrate on the
case where eigenvalues of
$A$ may occur anywhere on the complex plane, but no eigenvalues
of $A$ are on the imaginary axis and
no pairs of eigenvalues are located
symmetrically with respect to the imaginary axis. For such triplets
$(A,B,C)$ we show that there is an equivalent triplet
$(\tilde A,\tilde B,\tilde C)$ yielding the same solution
$u(x,t)$ to (1.1), where all eigenvalues of
$\tilde A$ have positive real parts. The corresponding
solutions $u(x,t)$ are then analytic
on the entire $xt$-plane and they are soliton solutions with any number of
poles in the corresponding transmission
coefficients and with any multiplicities of such poles. For such triplets
we also explicitly evaluate the corresponding transmission coefficients,
bound-state norming constants, and the corresponding
Jost solutions to (1.2).
We also consider the generalization of our results
to the matrix case, where the scalar quantity $u(x,t)$ in the NLS equation
is replaced by a matrix-valued function of $x$ and $t.$

Our paper is organized as follows. In Section 2 we present
the preliminary material by providing an outline of the method of
[3] and introduce exact solutions $u(x,t)$ to (1.1)
constructed via a triplet $(A,B,C).$
In Section 3 we exploit a certain symmetry in such exact solutions
and show that some (or all) eigenvalues of $A$ can be chosen
either on the right or on the left half complex plane
without changing $u(x,t).$
In Section 4 we show that such solutions are solitons with any number of
poles in the corresponding transmission
coefficients and with any multiplicities of such poles.
We also explicitly evaluate the corresponding transmission coefficients
and bound-state norming constants and the
Jost solutions to (1.2).
In Section 5 we show that the results of Sections 2-4 obtained for
the (scalar) NLS equation (1.1) remain valid
for the matrix NLS equation
$$iu_t+u_{xx}+2uu^\dagger u=0,\tag 1.3$$
where $u$ is now $m\times n$ matrix valued,
the dagger denotes the matrix adjoint (matrix transpose and complex conjugate),
and the associated Zakharov-Shabat system is given by
$$\displaystyle\frac{d \varphi(\lambda,x,t)}{dx}=\bmatrix -i\lambda I_m&u(x,t)\\
\noalign{\medskip}
-u(x,t)^\dagger&i\lambda I_n\endbmatrix \varphi(\lambda,x,t),\tag 1.4$$
with $I_n$ denoting the $n\times n$ identity matrix.
Finally, we conclude in Section 6 with an explicit example, which
was earlier studied as Example 7.2 of [3].
Since one of the eigenvalues of $A$ in that example has
a negative real part, we earlier conjectured
that it might be a nonsoliton solution. Using our current results, we verify in Section 6 that
it is a two-soliton solution and we explicitly evaluate the corresponding
transmission coefficients,
the bound-state norming constants, and the Jost solutions to (1.2).

Note that (1.1) and (1.3) are identical when $u$ is a scalar, and
similarly (1.2) and (1.4) are identical when $u$ is a scalar.
Throughout our paper we refer to (1.3) as the scalar
NLS equation when $u$ is a scalar and as the matrix NLS
equation when $u$ is matrix valued. This convenience enables
us to state all our results for the NLS equation
so that,
as shown in Section 5, they remain valid
whether the scalar case or the matrix case is considered.

\vskip 10 pt
\noindent {\bf 2. PRELIMINARIES}
\vskip 3 pt

In this section we establish our notation and provide the preliminaries for
certain exact solutions to the focusing NLS equation. Such exact solutions are expressed [3]
in terms of a triplet of matrices $(A,B,C).$

Consider any triplet
$(A,B,C),$ where
$A$ is a $p\times p$ (complex-valued) constant matrix,
$B$ is a $p\times 1$ (complex-valued) constant matrix, and $C$ is a
$1\times p$ (complex-valued) constant matrix. For short, we will refer to
such a triplet as a triplet of size $p.$ From such a triplet, construct
the auxiliary $p\times p$ matrices $Q$ and $N$ by solving
the respective Lyapunov equations
$$\cases QA+A^\dagger Q=C^\dagger C,\\
\noalign{\medskip}
AN+NA^\dagger=BB^\dagger.\endcases\tag 2.1$$
Note that if $(Q,N)$ satisfies the
system in (2.1), so does $(Q^\dagger,N^\dagger);$ hence, there is no loss of generality
in assuming that the solution matrices $Q$ and $N$ to (2.1) are selfadjoint.
Then, form the $p\times p$ matrix $F(x,t)$ and the scalar quantity $u(x,t)$ as
$$F(x,t):=e^{2A^\dagger x-4i(A^\dagger)^2t}+Q
\,e^{-2Ax-4iA^2t}N,\tag 2.2$$
$$u(x,t):=-2B^\dagger F(x,t)^{-1}
C^\dagger.\tag 2.3$$
Let us also define the $p\times p$ matrix $G(x,t)$ and the scalar quantity
$v(x,t)$ as
$$G(x,t):=e^{-2Ax-4iA^2t}+Ne^{2A^\dagger x-4i(A^\dagger)^2t}Q,\tag 2.4$$
$$v(x,t):=-2CG(x,t)^{-1}B.\tag 2.5$$

\noindent{\bf Theorem 2.1} {\it Given a triplet $(A,B,C)$ of size $p,$ let
$(Q,N)$ be a selfadjoint solution to the system in (2.1). Then $u(x,t)$ defined
as in (2.3) satisfies the NLS equation (1.3) 
at any point on the $xt$-plane where $F(x,t)$ is invertible.
Similarly, $v(x,t)$ defined in (2.5)
satisfies (1.3) at any point on the $xt$-plane where $G(x,t)$ is invertible.}

\noindent PROOF: Let us only give the proof for $v$ because the proof for $u$ is similar.
In fact, $v(x,t)$ turns into $u(-x,t)$ when replacing $(A,B,C)$ with
$(A^\dagger,C^\dagger,B^\dagger),$
which can also be used for the proof related to $u(x,t).$
Let us drop the arguments of the functions involved and write
(2.4) and its adjoint as
$$G=e^{-\beta}+Ne^{\beta^\dagger}Q,\qquad
G^\dagger=e^{-\beta^\dagger}+Qe^{\beta}N,\quad \tag 2.6$$
where we have defined
$$\beta:=2Ax+4iA^2t.\tag 2.7$$
Note that $v$ in (2.5) is well defined as long as $G^{-1}$ exists.
Taking appropriate partial derivatives, from (2.5) and (2.6) we obtain
$$iv_t+v_{xx}+2vv^\dagger v=-2C G^{-1}PG^{-1}B,\tag 2.8$$
$$P:=-iG_t-G_{xx}+2G_xG^{-1}G_x+8BB^\dagger (G^\dagger)^{-1}C^\dagger C.$$
 From (2.6) we get
$$G_t=-4iA^2 e^{-\beta}-4iN (A^\dagger)^2 e^{\beta^\dagger}Q,\quad
G_x=-2A e^{-\beta}+2N A^\dagger e^{\beta^\dagger}Q,\tag 2.9$$
$$G_{xx}=4A^2 e^{-\beta}+4N ( A^\dagger)^2 e^{\beta^\dagger}Q,\tag 2.10$$
$$e^{\beta^\dagger}QG^{-1}=(G^\dagger)^{-1}Qe^{\beta},\quad
e^{\beta}N(G^\dagger)^{-1}=G^{-1}Ne^{\beta^\dagger}.\tag 2.11$$
With the help of (2.1), (2.6), and (2.9)-(2.11), one can verify that
$P=0,$ and hence the right hand side of (2.8) is zero.  \qed

There are several questions that can be raised. For example, are the
Lyapunov equations given in (2.1) solvable; if they are solvable,
are they uniquely solvable? Are the matrices $F$ and $G$
defined in (2.2) and (2.4), respectively, invertible?
The answers to these questions are affirmative under appropriate restrictions
on the triplet $(A,B,C),$ as we will see.

Let us note that $u$ and $v$ defined in
(2.3) and (2.5), respectively, are analytic functions of $x$ and $t$
at any point on the $xt$-plane as long as the matrices
$F$ and $G,$ respectively,
are invertible at that point.
This is because the entries of those matrices
and hence also their determinants can
be written as sums of products of sine, cosine,
exponential, and polynomial functions of linear
combinations of $x$ and $t.$

Consider the scalar function $\Omega$ defined as
$$\Omega(x):=Ce^{-Ax}B.\tag 2.12$$
The right hand side is called a matrix realization of $\Omega$ in terms of
the triplet $(A,B,C).$
Without changing $\Omega(x),$
it is possible to increase the value of $p$ in the size of the triplet $(A,B,C)$
by padding the matrices $A,$ $B,$ $C$ with zeros or by modifying $A,$ $B,$ $C$
in some other fashion (cf. [8], Subsection 2.4).
Conversely, it might also be possible to reduce
the value of $p$ in the triplet $(A,B,C)$ so that the quantity
$\Omega(x)$ will remain unchanged.
The matrix realization in (2.12)
is said to be minimal if the value of $p$ in the triplet $(A,B,C)$ is the smallest
and yet $\Omega(x)$ remains unchanged by the choice of $p.$ The triplet $(A,B,C)$
is minimal if and only if [5]
the intersections of the kernels of $CA^j$ and of the kernels of
$B^\dagger(A^\dagger)^j$ for
$j=0,1,2,\dots$ are trivial; i.e.
$$\left\{\xi\in{\bold C}^p:CA^j\xi=0\ \text{for}\ j\ge0\right\}=\{0\}
=\left\{\eta\in{\bold C}^p:B^\dagger(A^\dagger)^j\eta=0\ \text{for}\ j\ge0\right\}.\tag 2.13$$
It is also known [5] that the triplet yielding a minimal realization
in (2.12) is unique up to
a similarity transformation $(A,B,C)\mapsto(EAE^{-1},EB,CE^{-1})$
for some unique matrix $E.$

The results in the next theorem are known [3], but they are
collected here in a summarized form
and a brief proof is included
for the benefit of the reader.

\noindent{\bf Theorem 2.2} {\it Assume that the triplet
$(A,B,C)$ of size $p$ corresponds to a minimal realization in (2.12) and that
the eigenvalues of $A$ all have positive real parts. Then:}
\item{(i)} {\it The Lyapunov equations in (2.1) are uniquely solvable.}
\item{(ii)} {\it The solutions $Q$ and $N$ are $p\times p$ selfadjoint matrices.}
\item{(iii)} {\it $Q$ and $N$ can be expressed in terms of the triplet
$(A,B,C)$ as}
$$Q=\int_0^\infty ds\,[Ce^{-As}]^\dagger [Ce^{-As}],\qquad
N=\int_0^\infty ds\,[e^{-As}B][e^{-As}B]^\dagger.\tag 2.14$$
\item{(iv)} {\it $Q$ and $N$ are invertible matrices.}
\item{(v)} {\it Any square submatrix of $Q$ containing
the (1,1)-entry or $(p,p)$-entry of $Q$ is invertible. Similarly,
any square submatrix of $N$ containing
the (1,1)-entry or $(p,p)$-entry of $N$ is invertible.}
\item{(vi)} {\it The quantities $F$ and $G$
defined in (2.2) and (2.4), respectively, are $p\times p$ matrices invertible
at every point on the $xt$-plane.}

\noindent PROOF: By introducing the parameter
$\alpha,$ let us write the first equation in (2.1) as
$$-Q(\alpha I-A)+(\alpha I+A^\dagger)Q=C^\dagger C,$$
or equivalently as
$$-(\alpha I +A^\dagger)^{-1}Q+Q(\alpha I-A)^{-1}=
(\alpha I+A^\dagger)^{-1}C^\dagger C(\alpha-A)^{-1},
\tag 2.15$$
where $I$ is the $p\times p$ identity matrix.
Since $A$ and $(-A^\dagger)$ have eigenvalues on the right and
left complex half planes, respectively, we can integrate (2.15)
along a simple
and positively oriented contour $\gamma$ lying on the
right half complex plane and enclosing
all eigenvalues of $A.$ Thus, we obtain $Q$ uniquely as
$$Q=\displaystyle\frac{1}{2\pi i}\int_\gamma d\alpha\,
(\alpha I+A^\dagger)^{-1}C^\dagger C(\alpha I-A)^{-1}.$$
Similarly, the solution to the second equation in (2.1) is unique and
is obtained as
$$N=\displaystyle\frac{1}{2\pi i}\int_\gamma d\alpha\,(\alpha I-A)^{-1}BB^\dagger
(\alpha I+A^\dagger)^{-1}.$$
Thus, (i) is proved. From (2.1) it is seen that $(Q^\dagger,N^\dagger)$ is a solution to
(2.1) whenever $(Q,N)$ is a solution and hence from the uniqueness of
the solution we obtain (ii). From (2.14) we see that
$$QA+A^\dagger Q=-\int_0^\infty
ds\,\displaystyle\frac{d}{ds}\left[e^{-A^\dagger s}C^\dagger Ce^{-As}\right]=-
e^{-A^\dagger s}C^\dagger Ce^{-As}\big|_{s=0}^\infty=C^\dagger C,$$
where we have used the fact that all eigenvalues of $A$ have positive real parts.
A similar argument for $N$ completes the proof of (iii).
 From their selfadjointness and positivity as seen from (2.14),
it follows that all eigenvalues of $Q$ and $N$
are nonnegative. Moreover, (2.13) implies that zero is not an eigenvalue of $Q$ and
$N$. Hence $Q$ and $N$ are invertible, proving (iv).
The positivity of all eigenvalues also implies (v).
The invertibility of $F$ follows from using Theorem~4.2 of [3]
in (2.2), and the proof of
invertibility for $G$ is similar. \qed

The results in the next theorem are useful in extracting
the scattering data for (1.2) from the corresponding
potential $u(x,t),$ which is also a solution to (1.1). For the benefit
of the reader we state such results in a summarized and unified form.
The proofs of these results are available in Theorems 3.1 and 3.3 of [3],
and hence they will not be given here.

\noindent{\bf Theorem 2.3} {\it Assume that the triplet
$(\tilde A,\tilde B,\tilde C)$ of size $p$ corresponds to a minimal realization in (2.12) and that
the eigenvalues of $\tilde A$ all have positive real parts. Further, assume that $\tilde A$ has
$m$ distinct eigenvalues $\alpha_1,\dots,\alpha_m$ and the multiplicity
of $\alpha_j$ is $n_j.$ Then:}
\item{(i)} {\it There exists
a unique triplet $(A,B,C),$ where
$A$ is in a Jordan canonical form with each Jordan block
containing a distinct eigenvalue and having $-1$ in the superdiagonal entries, and
the entries of $B$ consist of zeros and ones. More specifically, we have}
$$A=\bmatrix A_1&
0&\dots&0\\
0& A_2&\dots&0\\
\vdots&\vdots &\ddots&\vdots\\
0&0&\dots&A_m\endbmatrix,
\quad
B=\bmatrix
B_1\\
B_2\\
\vdots\\
B_m\endbmatrix,\quad
C=\bmatrix C_1&C_2& \dots
&
C_m\endbmatrix,$$
$$A_j:=\bmatrix \alpha_j&
-1&0&\dots&0\\
0& \alpha_j& -1&\dots&0\\
0&0&\alpha_j& \dots&0\\
\vdots&\vdots &\vdots &\ddots&\vdots\\
0&0&0&\dots&\alpha_j\endbmatrix,
\quad
B_j:=\bmatrix
0\\
0\\
0\\
\vdots\\
1\endbmatrix,
\quad C_j:=\bmatrix c_{j(n_j-1)}& \dots
&
c_{j 1}&
c_{j 0}\endbmatrix,$$
{\it where $A_j$ has size $n_j\times n_j,$ $B_j$ has size $n_j\times 1,$
$C_j$ has size $1\times n_j,$  and
the constants $c_{j(n_j-1)}$ are nonzero.}
\item{(ii)} {\it The triplet $(A,B,C)$ can be constructed from $(\tilde A,\tilde B,\tilde C)$
via}
$$\tilde A=M A M^{-1},\quad \tilde B=MSB,\quad C=\tilde CMS,\tag 2.16$$
{\it where $M$ is a matrix whose columns are formed from the generalized
eigenvectors of $(-\tilde A)$ and
$S$ is an upper triangular Toeplitz matrix
commuting with $A,$ is uniquely determined by $M$ and $\tilde B,$ and has the form}
$$S=\bmatrix S_1&0&\dots&0\\
0&S_2&\dots&0\\
0&0& \dots&0\\
\vdots&\vdots &\ddots&\vdots\\
0&0&\dots&S_{m+n}\endbmatrix,
\quad S_j:=
\bmatrix \theta_{jn_j}&\theta_{j(n_j-1)}&\dots&\theta_{j1}\\
0&\theta_{jn_j}&\dots&\theta_{j2}\\
0&0& \dots&\theta_{j3}\\
\vdots&\vdots &\ddots&\vdots\\
0&0&\dots&\theta_{jn_j}\endbmatrix,$$
{\it for some constants $\theta_{js}.$}
\item{(iii)} {\it The triplets
$(A,B,C)$ and $(\tilde A,\tilde B,\tilde C)$
yield the same solution $u(x,t)$ to
(1.1).}
\item{(iv)} {\it The complex constants
$(i\alpha_j)$ correspond
to the bound-state poles on the upper half complex plane
of the
transmission coefficients $T_{\text l}$ and $T_{\text r}$
appearing in (4.4)-(4.7) of Section~4.}
\item{(v)} {\it  For each $j,$ the complex constants
$c_{js}$ for $s=0,1,\dots,(n_j-1)$ appearing in $C$
correspond to the bound-state norming constants associated with the bound-state pole
$(i\alpha_j)$ of the transmission coefficients.}

\vskip 10 pt
\noindent {\bf 3. THE SCALAR NLS EQUATION}
\vskip 3 pt

Recall that we refer to (1.3) as the scalar NLS equation
when $u$ is a scalar quantity and
as the matrix NLS equation when
$u$ is a matrix quantity.
In this section we exploit a certain symmetry in (2.2)
and show that without changing the value of the scalar
$u(x,t)$ in (2.3) it is possible to transform
the triplet $(A,B,C)$ in such a way that some or all eigenvalues of $A$ can be reflected
from the right half complex plane to the left half complex plane.
The same result holds for the scalar
$v(x,t)$ given in (2.5); namely,
it remains unchanged when the triplet $(A,B,C)$
is transformed so that some or all eigenvalues
of $A$ are reflected with respect to
the imaginary axis on the complex plane.
Later, we will see that these results remain valid
also for the matrix NLS equation when $u$ and $v$ are matrix valued.

For repeated eigenvalues of $A,$ the aforementioned transformation
must be applied to all the multiplicities in such a way that
after the transformation we should not have any eigenvalue
pairs symmetrically located with respect to
the imaginary axis of the complex plane. As mentioned in Section 1,
eigenvalue pairs symmetrically located with respect to
the imaginary axis cannot yield soliton solutions to (1.1).
For such pairs, it is already known [9] that (2.1)
is not uniquely solvable.

Let us write (2.2) as
$$F(x,t)=Q\left[e^{-2Ax-4iA^2t}+Q^{-1}e^{2A^\dagger x-4i(A^\dagger)^2t}N^{-1}\right]
N.\tag 3.1$$
Comparing (2.2) and (3.1), we next prove that $u(x,t)$ appearing in (2.3)
remains invariant under the transformation
$$(A,B,C,Q,N)\mapsto (-A^\dagger,-N^{-1}B,-CQ^{-1},-Q^{-1},-N^{-1}),$$
where all the eigenvalues of $A$ are reflected with respect
to the imaginary axis on the complex plane as a result of $A\mapsto (-A^\dagger).$

\noindent{\bf Theorem 3.1} {\it Assume that the triplet $(A,B,C)$
corresponds to a minimal realization in (2.12) and that all eigenvalues of $A$
have positive real parts. Consider the transformation}
$$(A,B,C,Q,N,F,G,u,v)\mapsto(\tilde A,\tilde B,\tilde C,
\tilde Q,\tilde N,\tilde F,\tilde G,\tilde u,\tilde v),\tag 3.2$$
{\it where $(Q,N)$ corresponds to the unique
solution to the Lyapunov system in (2.1),
the quantities in $(F,G,u,v)$ are as
in (2.2)-(2.5),}
$$\tilde A=-A^\dagger,\quad \tilde B=-N^{-1}B,\quad
\tilde C=-CQ^{-1},\quad \tilde Q=-Q^{-1},\quad \tilde N=-N^{-1},\tag 3.3$$
{\it and $(\tilde F,\tilde G,\tilde u,\tilde v)$
is as in (2.2)-(2.5)
but by using $(\tilde A,\tilde B,\tilde C,\tilde Q,\tilde N)$ instead of $(A,B,C,Q,N)$
on the right hand sides. We then have the following:}
\item{(a)} {\it The quantities $F$ and $G$ are transformed as}
$$\tilde F=Q^{-1}FN^{-1},\quad \tilde G=N^{-1}GQ^{-1}.\tag 3.4$$
\item{(b)} {\it $\tilde Q$ and $\tilde N$ satisfy the respective
Lyapunov equations}
$$\cases \tilde Q\tilde A+\tilde A^\dagger \tilde Q=\tilde C^\dagger \tilde C,\\
\noalign{\medskip}
\tilde A\tilde N+\tilde N\tilde A^\dagger=\tilde B\tilde B^\dagger.\endcases\tag 3.5$$
\item{(c)} {\it The matrices $\tilde Q$ and $\tilde N$ are selfadjoint and invertible.}
\item{(d)} {\it The matrices $\tilde F$ and $\tilde G$ are invertible
at every point on the $xt$-plane.}
\item{(e)} {\it $\tilde u(x,t)=u(x,t)$ and $\tilde v(x,t)=v(x,t).$}

\noindent PROOF:
Using (3.3) in (2.2) and (2.4) we get (a).
Using (2.1) and (3.3), it can directly be verified that
(3.5) is satisfied, proving (b). Using (ii) and (iv) of Theorem 2.2 in (3.3),
it follows that (c) holds. The invertibility in (d)
follows from (3.4) by using the
invertibility of $Q$ and $N$ stated in (c)
and the invertibility of $F$ and $G$ stated in (vi) of Theorem 2.2.
Finally, (e) can be proved
by using (2.3) and (2.5)
with the help of (3.3), the selfadjointness
of $Q$ and $N,$ and (3.4).
\qed

Next, we show that even if we reflect some of eigenvalues of
$A$ from the right to the left half complex plane, Theorem 3.1 remains valid
by choosing the transformation in (3.2) appropriately. For this purpose,
let us again start with a triplet
$(A,B,C)$ of size $p$ and corresponding to a minimal realization in (2.12),
where the eigenvalues of $A$ all have positive real parts. Without loss of any generality,
let us partition $A,$ $B,$ $C$
as
$$A=\bmatrix A_1&
0\\
\noalign{\medskip}
0&A_2\endbmatrix,\quad B=\bmatrix B_1\\
\noalign{\medskip}
B_2\endbmatrix,\quad C=\bmatrix C_1&C_2\endbmatrix,\tag 3.6$$
so that the $q\times q$ block diagonal matrix $A_1$ contains the eigenvalues
that will remain unchanged and
$A_2$ contains the eigenvalues
that will be reflected with respect to the imaginary axis
on the complex plane, the submatrices $B_1$ and $C_1$ have sizes
$q\times 1$ and $1\times q,$ respectively, and hence
$A_2,$ $B_2,$ $C_2$ have sizes
$(p-q)\times(p-q),$ $(p-q)\times 1,$ $1\times (p-q),$ respectively,
for some integer $q$ not exceeding $p.$
Let us write the corresponding respective solutions to (2.1)
as
$$Q=\bmatrix Q_1&Q_2\\
\noalign{\medskip}
Q_3&Q_4\endbmatrix,\quad N=\bmatrix N_1&N_2\\
\noalign{\medskip}
N_3&N_4\endbmatrix,\tag 3.7$$
where $Q_1$ and $N_1$ have sizes $q\times q,$ $Q_4$ and $N_4$ have sizes $(p-q)\times (p-q),$
etc.
Note that because of the selfadjointness of $Q$ and $N$ stated in Theorem 2.2, we have
$$Q_1^\dagger=Q_1,\quad
Q_2^\dagger=Q_3,\quad Q_4^\dagger=Q_4,\quad
N_1^\dagger=N_1,\quad
N_2^\dagger=N_3,\quad N_4^\dagger=N_4.\tag 3.8$$
Furthermore, from Theorem 2.2 (v) it follows that $Q_1,$ $Q_4$, $N_1,$ and
$N_4$ are all invertible.

Let us clarify our notational choice in (3.6) and emphasize that
the partitioning in (3.6) is not the same partitioning used in (i)
of Theorem 2.3.

\noindent{\bf Theorem 3.2} {\it Assume that the triplet $(A,B,C)$ partitioned as in (3.6)
corresponds to a minimal realization in (2.12) and that all eigenvalues of $A$
have positive real parts. Consider the transformation (3.2)
with $(\tilde A,\tilde B,\tilde C)$ having similar
block representations as in (3.6), $(Q,N)$ as in (3.7)
corresponding to the unique solution to
the Lyapunov system in (2.2),}
$$\tilde A_1=A_1,\quad \tilde A_2=-A_2^\dagger,\quad
\tilde B_1=B_1-N_2N_4^{-1}B_2,\quad
\tilde B_2=-N_4^{-1}B_2,\tag 3.9$$
$$\tilde C_1=C_1-C_2Q_4^{-1}Q_3,\quad
\tilde C_2=-C_2Q_4^{-1},\tag 3.10$$
{\it and $(\tilde Q,\tilde N)$ given as}
$$\tilde Q_1=Q_1-Q_2Q_4^{-1}Q_3,\quad
\tilde Q_2=-Q_2Q_4^{-1},\quad
\tilde Q_3=-Q_4^{-1}Q_3,\quad
\tilde Q_4=-Q_4^{-1},\tag 3.11$$
$$\tilde N_1=N_1-N_2N_4^{-1}N_3,\quad \tilde N_2=-N_2N_4^{-1},\quad
\tilde N_3=-N_4^{-1}N_3,\quad
\tilde N_4=-N_4^{-1},\tag 3.12$$
{\it and $(\tilde F,\tilde G,\tilde u,\tilde v)$
as in (2.2)-(2.5)
but by using $(\tilde A,\tilde B,\tilde C,\tilde Q,\tilde N)$ instead of $(A,B,C,Q,N)$
on the right hand sides. We then have the following:}

\item{(a)} {\it The quantities $F$ and $G$ are transformed
according to}
$$\tilde F=\bmatrix I&-Q_2Q_4^{-1}\\
\noalign{\medskip}
0&-Q_4^{-1}\endbmatrix F
\bmatrix I&0\\
\noalign{\medskip}
-N_4^{-1}N_3&-N_4^{-1}\endbmatrix,\tag 3.13$$
$$\tilde G=\bmatrix I&-N_2N_4^{-1}\\
\noalign{\medskip}
0&-N_4^{-1}\endbmatrix G
\bmatrix I&0\\
\noalign{\medskip}
-Q_4^{-1}Q_3&-Q_4^{-1}\endbmatrix.\tag 3.14$$

\item{(b)} {\it $\tilde Q$ and $\tilde N$ satisfy the respective
Lyapunov equations in (3.5).}
\item{(c)} {\it The matrices $\tilde Q$ and $\tilde N$ are selfadjoint and invertible.}
\item{(d)} {\it The matrices $\tilde F$ and $\tilde G$ are invertible
at every point on the $xt$-plane.}
\item{(e)} {\it $\tilde u(x,t)=u(x,t)$ and $\tilde v(x,t)=v(x,t).$}

\noindent PROOF: Let us use $I$ to denote the identity matrix not necessarily
having the same dimension in every appearance in the proof, but that dimension
will be apparent to the reader. Note that (3.13) and (3.14) can be verified
by using (3.8)-(3.12) in (2.2) and (2.4), which proves (a).
The proof of (b) is by direct substitution in
(3.5) and by using (2.1) and (3.6)-(3.12)
and by noting that
$$\tilde B=\bmatrix I&-N_2N_4^{-1}\\
\noalign{\medskip}
0&-N_4^{-1}\endbmatrix B=\bmatrix
I&-N_2\\
\noalign{\medskip}
0&-N_4\endbmatrix^{-1}B,\tag 3.15$$
$$\tilde C=C\bmatrix I&0\\
\noalign{\medskip}
-Q_4^{-1}Q_3&-Q_4^{-1}\endbmatrix=
C\bmatrix
I&0\\
\noalign{\medskip}
-Q_3&-Q_4\endbmatrix^{-1},\tag 3.16$$
where the invertibility of $Q_4$ and $N_4$ is also used, which follows from Theorem 2.2 (v).
The selfadjointness of $\tilde Q$ and $\tilde N$ follows from (3.7),
(3.8), (3.11), and (3.12). The invertibility
of $\tilde Q$ and $\tilde N$ can be seen from (3.11) and (3.12) by writing
$$\tilde Q=\bmatrix Q_1&Q_2\\
\noalign{\medskip}
0&I\endbmatrix\bmatrix I&0\\
\noalign{\medskip}
-Q_4^{-1}Q_3&-Q_4^{-1}\endbmatrix=
\bmatrix Q_1^{-1}&-Q_1^{-1}Q_2\\
\noalign{\medskip}
0&I\endbmatrix^{-1}\bmatrix I&0\\
\noalign{\medskip}
-Q_3&-Q_4\endbmatrix^{-1},\tag 3.17$$
and a similar expression for $\tilde N$
and the fact that $Q_1,$ $Q_4$, $N_1,$
$N_4$ are all invertible. Thus, (c) is proved.
The invertibility in (d)
follows from (3.13) and (3.14) by using the
invertibility of $Q_4$ and $N_4$ stated in (v) of Theorem 2.2
and the invertibility of $F$ and $G$ stated in (vi) of Theorem 2.2.
Finally, we evaluate
$\tilde F(x,t)$ by using $(\tilde A,\tilde B,\tilde C,\tilde Q,\tilde N)$ instead of $(A,B,C,Q,N)$
on the right hand side of (2.2) and evaluate
$\tilde u(x,t)$ by using $(\tilde B,\tilde F,\tilde C)$ instead of $(B,F,C)$
on the right hand side of (2.3). With the help of (3.7)-(3.17), it is straightforward to show that
$\tilde u(x,t)$ simplifies to
$u(x,t)$;
similarly, $\tilde v(x,t)$ simplifies to
$v(x,t),$ completing the proof of (e). \qed

Finally in this section, we show that $u$ and $v$ defined in (2.3) and (2.5),
respectively, can be transformed into each other
by transforming the triplet $(A,B,C)$ in a particular way.

\noindent{\bf Theorem 3.3} {\it Assume that the triplet $(A,B,C)$
corresponds to a minimal realization in (2.12) and that all eigenvalues of $A$
have positive real parts. Consider the transformation (3.2)}
{\it where $(Q,N)$ corresponds to the unique
solution to the Lyapunov system in (2.1) and $(F,G,u,v)$ is as
in (2.2)-(2.5),}
$$\tilde A=A,\quad \tilde B=Q^{-1}C^\dagger,\quad
\tilde C=B^\dagger N^{-1},\quad \tilde Q=N^{-1},\quad \tilde N=Q^{-1},\tag 3.18$$
{\it and $(\tilde F,\tilde G,\tilde u,\tilde v)$
is as in (2.2)-(2.5)
but by using $(\tilde A,\tilde B,\tilde C,\tilde Q,\tilde N)$ instead of $(A,B,C,Q,N)$
on the right hand sides. We then have the following:}
\item{(a)} {\it $\tilde Q$ and $\tilde N$ satisfy the respective
Lyapunov equations given in (3.5).}
\item{(b)} {\it The matrices $\tilde Q$ and $\tilde N$ are selfadjoint and invertible.}
\item{(c)} {\it $\tilde u(x,t)=v(x,t)$ and $\tilde v(x,t)=u(x,t).$}

\noindent PROOF: Using (2.1) and (3.18), it can directly be verified that
(3.5) is satisfied. The selfadjointness and invertibility
of $\tilde Q$ and $\tilde N$ follow from (3.18) because
$Q$ and $N$ have those properties, as stated in Theorem 2.2 (ii) and (iv);
hence, (b) holds. Finally, (c) is proved
by direct substitution
in (2.3) and (2.5) with the help of the selfadjointness
of $Q$ and $N.$ \qed

\vskip 10 pt
\noindent {\bf 4. THE SCATTERING COEFFICIENTS AND JOST SOLUTIONS}
\vskip 3 pt

In this section, we evaluate the Jost solutions and
the corresponding scattering coefficients for the Zakharov-Shabat system (1.2)
associated with the NLS equation (1.3) when the potential is given by
(2.3) or (2.5).
In analyzing (1.2), we will use the notation
$\varphi=\bmatrix \varphi_1\\
\varphi_2\endbmatrix,$ where the subscripts $1$ and $2$ denote the first and second components.
If $\varphi(\lambda,x,t)$ and $\bar \varphi(\lambda,x,t)$ are any two solutions
to (1.2), their Wronskian $[\varphi;\bar \varphi]$ is independent of $x,$ where we have defined
$$[\varphi;\bar \varphi]:=\varphi(\lambda^*,x,t)^\dagger
\bar \varphi(\lambda,x,t).\tag 4.1$$
We stress that
an overbar does not indicate
complex conjugation.

Recall [13,15] that the Jost solutions $\psi,$
$\bar \psi,$
$\phi,$
and $\bar \phi$
are defined
as those vector solutions
to (1.2) with asymptotics
$$\psi(\lambda,x,t)=\bmatrix 0\\
\noalign{\medskip}
e^{i\lambda x}\endbmatrix [1+o(1)],
\quad
\bar\psi(\lambda,x,t)=\bmatrix e^{-i\lambda x}\\
\noalign{\medskip}
0\endbmatrix [1+o(1)],
\qquad x\to+\infty,\tag 4.2$$
$$\phi(\lambda,x,t)=\bmatrix e^{-i\lambda x}\\
\noalign{\medskip}
0\endbmatrix [1+o(1)],
\quad
\bar\phi(\lambda,x,t)=\bmatrix 0\\
\noalign{\medskip}
e^{i\lambda x}\endbmatrix [1+o(1)],
\qquad x\to-\infty.\tag 4.3$$
When $u(\cdot,t)$ is integrable
for each fixed $t,$ the four Jost solutions exist
and their asymptotics at the other end of the real axis
yield the scattering coefficients $R,$ $L,$ $T_{\text l},$ $T_{\text r}$
via
$$\psi(\lambda,x,t)=\bmatrix LT_{\text l}^{-1}e^{-i\lambda x}\\
\noalign{\medskip}
T_{\text l}^{-1}e^{i\lambda x}\endbmatrix [1+o(1)],\qquad x\to-\infty,\tag 4.4$$
$$
\bar\psi(\lambda,x,t)=\bmatrix (T_{\text r}^\dagger)^{-1}e^{-i\lambda x}\\
\noalign{\medskip}
-L^\dagger(T_{\text r}^\dagger)^{-1} e^{i\lambda x}\endbmatrix [1+o(1)],
\qquad x\to-\infty,\tag 4.5$$
$$\phi(\lambda,x,t)=\bmatrix T_{\text r}^{-1}e^{-i\lambda x}\\
\noalign{\medskip}
RT_{\text r}^{-1}e^{i\lambda x}\endbmatrix [1+o(1)],\qquad x\to+\infty,\tag 4.6$$
$$
\bar\phi(\lambda,x,t)=\bmatrix - R^\dagger (T_{\text l}^\dagger)^{-1}e^{-i\lambda x}\\
\noalign{\medskip}
(T_{\text l}^\dagger)^{-1}e^{i\lambda x}\endbmatrix [1+o(1)],
\qquad x\to+\infty.\tag 4.7$$
The scattering coefficients can equivalently be obtained with the help
of the Wronskian defined in (4.1); in fact, we have
$$LT_{\text l}^{-1}=(RT_{\text r}^{-1})^\dagger=[\phi;\psi],\quad
T_{\text l}^{-1}=[\bar\phi;\psi],\quad
T_{\text r}^{-1}=[\bar\psi;\phi].\tag 4.8$$
For the Zakharov-Shabat system (1.2) we have $T_{\text l}=T_{\text r}$ if
the scalar potential $u(x,t)$
vanishes as $x\to\pm\infty.$ However, we will retain
the separate notations for $T_l$ and $T_r$ for
a subsequent generalization to the matrix case, where $T_l$ and $T_r$ may differ.

The Jost solutions $\psi$ and $\bar\psi$ and the potential
$u(x,t)$ are recovered [13,15] as
$$\psi(\lambda,x,t)=\bmatrix
0\\
\noalign{\medskip}
e^{i\lambda x}\endbmatrix+\int_x^\infty dy\,K(x,y,t)\,e^{i\lambda y},\tag 4.9$$
$$\bar\psi(\lambda,x,t)=\bmatrix
e^{-i\lambda x}\\
\noalign{\medskip}
0
\endbmatrix+\int_x^\infty dy\,\bar K(x,y,t)\,e^{-i\lambda y},\tag 4.10$$
$$u(x,t)=-2\bmatrix 1&0\endbmatrix K(x,x,t)=2\,\bar K(x,x,t)^\dagger \bmatrix 0\\
1\endbmatrix,\tag 4.11$$
 from the solutions to the Marchenko integral equations
$$\bar K(x,y,t)+\bmatrix 0\\
\noalign{\medskip}
\Omega_{\text l}(x+y,t)\endbmatrix+\int_x^\infty dy\,K(x,z,t)\,\Omega_{\text l}(z+y,t)=\bmatrix 0\\
\noalign{\medskip}
0\endbmatrix,\qquad y>x,\tag 4.12$$
$$K(x,y,t)-\bmatrix \Omega_{\text l}(x+y,t)^\dagger\\
\noalign{\medskip}
0\endbmatrix-\int_x^\infty dy\,\bar K(x,z,t)\,\Omega_{\text l}(z+y,t)^\dagger=\bmatrix 0\\
\noalign{\medskip}
0\endbmatrix,\qquad  y>x.\tag 4.13$$
The Jost solutions $\psi$ and $\bar\psi$ corresponding to $u(x,t)$ given in
(2.3) with $(A,B,C)=(A_{\text l},B_{\text l},C_{\text l})$
can be evaluated by solving (4.12) and (4.13) with
$$\Omega_{\text l}(y,t)=C_{\text l}e^{-A_{\text l}y-4iA_{\text l}^2 t}B_{\text l}.\tag 4.14$$
Since
$$\Omega_{\text l}(x+y,t)=C_{\text l}e^{-A_{\text l}x}
e^{-A_{\text l}y-4iA_{\text l}^2 t}B_{\text l},$$
the Marchenko equations in (4.12) and (4.13) have integral
kernels separable in $z$ and $y$ and hence
they can be solved explicitly by algebraic methods, yielding
$$\psi(\lambda,x,t)=\bmatrix iB_{\text l}^\dagger
F_{\text l}(x,t)^{-1}(\lambda I+iA_{\text l}^\dagger)^{-1}
e^{i\lambda x}
C_{\text l}^\dagger\\
\noalign{\medskip}
e^{i\lambda x}-iC_{\text l}[F_{\text l}(x,t)^\dagger]^{-1}N_{\text l}
e^{-2A_{\text l}^\dagger x+4i(A_{\text l}^\dagger)^2t}
(\lambda I+iA_{\text l}^\dagger)^{-1}e^{i\lambda x}
C_{\text l}^\dagger
\endbmatrix,\tag 4.15$$
$$\bar\psi(\lambda,x,t)=\bmatrix
e^{-i\lambda x}+iB_{\text l}^\dagger
e^{-2A_{\text l}^\dagger x+4i(A_{\text l}^\dagger)^2t}
Q_{\text l}
[F_{\text l}(x,t)^\dagger]^{-1}
(\lambda I-iA_{\text l})^{-1}e^{-i\lambda x}B_{\text l}\\
\noalign{\medskip}
iC_{\text l}
[F_{\text l}(x,t)^\dagger]^{-1}
(\lambda I-iA_{\text l})^{-1}e^{-i\lambda x}B_{\text l}
\endbmatrix.\tag 4.16$$

Similarly, the Jost solutions $\phi$ and $\bar\phi$
and the potential $v(x,t),$ in the form given in (2.5),
are recovered as
$$\phi(\lambda,x,t)=\bmatrix
e^{-i\lambda x}\\
\noalign{\medskip}
0
\endbmatrix+\int_{-\infty}^x dy\,M(x,y,t)\,e^{-i\lambda y},\tag 4.17$$
$$\bar\phi(\lambda,x,t)=\bmatrix
0\\
\noalign{\medskip}
e^{i\lambda x}
\endbmatrix+\int_{-\infty}^x dy\,\bar M(x,y,t)\,e^{i\lambda y},\tag 4.18$$
$$v(x,t)=2\bmatrix 1&0\endbmatrix \bar M(x,x,t)=-2\,M(x,x,t)^\dagger \bmatrix 0\\
1\endbmatrix,\tag 4.19$$
 from the solutions to
the Marchenko integral equations
$$\bar M(x,y,t)+\bmatrix \Omega_{\text r}(x+y,t)\\
\noalign{\medskip}
0\endbmatrix+\int_{-\infty}^x dy\,M(x,z,t)\,\Omega_{\text r}(z+y,t)=\bmatrix 0\\
\noalign{\medskip}
0\endbmatrix,\qquad y<x,\tag 4.20$$
$$M(x,y,t)-\bmatrix 0\\
\noalign{\medskip}
\Omega_{\text r}(x+y,t)^\dagger\endbmatrix
-\int_{-\infty}^x dy\,\bar M(x,z,t)\,\Omega_{\text r}(z+y,t)^\dagger=\bmatrix 0\\
\noalign{\medskip}
0\endbmatrix,\qquad  y<x.\tag 4.21$$
Corresponding to the potential $v(x,t)$ given in
(2.5) with $(A,B,C)=(A_{\text r},B_{\text r},C_{\text r})$
we have
$$\Omega_{\text r}(y,t)=C_{\text r}e^{A_{\text r}y+4iA_{\text r}^2 t}B_{\text r},\tag 4.22$$
which yields an integral kernel separable in $z$ and $y$
in (4.20) and (4.21) and hence allows us to solve those integral
equations explicitly by algebraic methods, yielding
$$\phi(\lambda,x,t)=\bmatrix e^{-i\lambda x}-i
C_{\text r}
G_{\text r}(x,t)^{-1}
N_{\text r}
e^{2A_{\text r}^\dagger x-4i(A_{\text r}^\dagger)^2t}
(\lambda I+iA_{\text r}^\dagger)^{-1}e^{-i\lambda x}C_{\text r}^\dagger\\
\noalign{\medskip}
iB_{\text r}^\dagger[G_{\text r}(x,t)^\dagger]^{-1}
(\lambda I+iA_{\text r}^\dagger)^{-1}e^{-i\lambda x}C_{\text r}^\dagger\endbmatrix,
\tag 4.23$$
$$\bar\phi(\lambda,x,t)=\bmatrix
iC_{\text r}
G_{\text r}(x,t)^{-1}
(\lambda I-iA_{\text r})^{-1}
e^{i\lambda x}B_{\text r}\\
\noalign{\medskip}
e^{i\lambda x}+
iB_{\text r}^\dagger
[G_{\text r}(x,t)^\dagger]^{-1}
Q_{\text r}
e^{2A_{\text r} x+4iA_{\text r}^2t}
(\lambda I-iA_{\text r})^{-1}
e^{i\lambda x}
B_{\text r}
\endbmatrix.\tag 4.24$$

\noindent{\bf Proposition 4.1} {\it Assume that the triplet $(A,B,C)$
corresponds to a minimal realization in (2.12) and that all eigenvalues of $A$
have positive real parts. Let $(Q,N)$ correspond to the unique
solution to the Lyapunov system (2.1) and let $(F,G)$ be as
in (2.2) and (2.4). We then have:}
\item{(a)} {\it $F(x,t)^{-1}\to 0$
and $G(x,t)^{-1}\to 0$
as $x\to\pm\infty.$}
\item{(b)} {\it $
e^{-2A^\dagger x+4i(A^\dagger)^2t}Q
[F(x,t)^\dagger]^{-1}\to N^{-1}$ as $x\to-\infty$.}
\item{(c)} {\it $[F(x,t)^\dagger]^{-1}N
e^{-2A^\dagger x+4i(A^\dagger)^2t}\to Q^{-1}$ as $x\to-\infty$.}
\item{(d)} {\it $[G(x,t)^\dagger]^{-1}Q
e^{2A x+4iA^2t}\to N^{-1}$ as $x\to+\infty$.}
\item{(e)} {\it $G(x,t)^{-1}N
e^{2A^\dagger x-4i(A^\dagger)^2t}\to Q^{-1}$ as $x\to+\infty$.}

\noindent PROOF: The proofs involving
$G$ are obtained by using (2.6)
the same way as in the proofs for $F,$ and hence we will only give the proofs for $F.$
Using (2.2) and (2.7) we see that
$$F=e^{\beta^\dagger}+Qe^{-\beta}N,\quad F^\dagger=e^{\beta}+Ne^{-\beta^\dagger}Q,$$
and hence
$$F=e^{\beta^\dagger}[I+e^{-\beta^\dagger}Qe^{-\beta}N],
\quad Q^{-1}FN^{-1}e^{\beta}=I+Q^{-1}e^{\beta^\dagger}N^{-1}e^{\beta},\tag 4.25$$
$$F^\dagger Q^{-1}e^{\beta^\dagger}N^{-1}=[I+e^{\beta}Q^{-1}e^{\beta^\dagger}N^{-1}],\quad
e^{\beta^\dagger}N^{-1}F^\dagger =[I+e^{\beta^\dagger}N^{-1}e^{\beta}Q^{-1}]Q.\tag 4.26$$
Forming the inverses from (4.25) and (4.26) we obtain
$$F^{-1}=[I+e^{-\beta^\dagger}Qe^{-\beta}N]^{-1}e^{-\beta^\dagger},\quad
F^{-1}=N^{-1}e^{\beta}[I+Q^{-1}e^{\beta^\dagger}N^{-1}e^{\beta}]^{-1}Q^{-1},\tag 4.27$$
$$e^{-\beta^\dagger}Q(F^\dagger)^{-1}=N^{-1}[I+e^{\beta}Q^{-1}e^{\beta^\dagger}N^{-1}]^{-1},
\tag 4.28$$
$$(F^\dagger)^{-1}Ne^{-\beta^\dagger}=Q^{-1}[I+e^{\beta^\dagger}N^{-1}e^{\beta}Q^{-1}]^{-1}.
\tag 4.29$$
By letting $x\to\pm\infty$ in (4.27)
we prove (a). By letting $x\to-\infty$ in (4.28) and (4.29) we prove
(b) and (c), respectively.
The remaining proofs involving $G$ are obtained
in a similar manner. \qed

The following result is known [5], but it will be useful later and hence we
provide a brief proof.

\noindent{\bf Proposition 4.2} {\it
For any triplet of matrices $(A,B,C)$ with sizes
$p\times p,$ $p\times n,$ and $n\times p,$
respectively,
we have}
$$[I_n+C(\lambda I_p-A)^{-1}B]^{-1}=I_n-C(\lambda I_p-A+BC)^{-1}B.\tag 4.30$$

\noindent PROOF: Using the identity
$$BC=(\lambda I_p-A+BC)-(\lambda I_p-A),$$
one can verify that the product of
one side of (4.30) and its inverse yields $I_n.$ \qed

 For convenience we also give a short proof of the following already known result on
Schur complements [9].

\noindent{\bf Proposition 4.3} {\it For any matrices $U$ and $V$ with sizes
$n\times p$ and $p\times n,$ respectively, we have the determinant identity}
$$\det(I_n+UV)=\det(I_p+VU).\tag 4.31$$

\noindent PROOF: Let us use the decompositions
$$\bmatrix I_p+VU&0_{pn}\\
\noalign{\medskip}
0_{np}&I_n\endbmatrix=\bmatrix I_p&-V\\
\noalign{\medskip}
0_{np}&I_n\endbmatrix \bmatrix I_p&V\\
\noalign{\medskip}
-U&I_n\endbmatrix \bmatrix I_p&0_{pn}\\
\noalign{\medskip}
U&I_n\endbmatrix,\tag 4.32$$
$$\bmatrix I_p&0_{pn}\\
\noalign{\medskip}
0_{np}&I_n+UV\endbmatrix=\bmatrix I_p&0_{pn}\\
\noalign{\medskip}
U&I_n\endbmatrix \bmatrix I_p&V\\
\noalign{\medskip}
-U&I_n\endbmatrix \bmatrix I_p&-V\\
\noalign{\medskip}
0_{np}&I_n\endbmatrix,\tag 4.33$$
where $0_{jk}$ denotes the zero matrix of size $j\times k.$
Note that the right hand sides in (4.32) and (4.33)
have the same determinant, and by equating
the determinants of the left hand sides we get (4.31). \qed

\noindent{\bf Theorem 4.4} {\it Assume that the triplet $(A,B,C)$
corresponds to a minimal realization in (2.12) and that all eigenvalues of $A$
have positive real parts. Let $(Q,N)$ correspond to the unique
solution to the Lyapunov system in (2.1), $(F,u)$ be as in
(2.2) and (2.3),
and the scattering coefficients be defined as in
(4.4)-(4.7). We then have:}

\item{(a)} {\it $u(x,t)\to 0$ as $x\to\pm\infty.$}
\item{(b)} {\it For $\lambda\in\bold R,$ the transmission coefficients $T_{\text l}$ and
$T_{\text r}$ appearing in (4.4) and (4.6),
respectively, and their inverses are given by}
$$T_{\text l}^{-1}=1-iCQ^{-1}
(\lambda I+iA^\dagger)^{-1}C^\dagger,
\tag 4.34$$
$$T_{\text l}=1+iC(\lambda I-iA)^{-1}Q^{-1}C^\dagger,
\tag 4.35$$
$$T_{\text r}^{-1}=1-iB^\dagger
(\lambda I+iA^\dagger)^{-1} N^{-1}B,
\tag 4.36$$
$$T_{\text r}=1+iB^\dagger N^{-1}(\lambda I-iA)^{-1}B,
\tag 4.37$$
{\it and hence they are functions of $\lambda$ alone and do not
depend on $t.$}
\item{(c)} {\it The reflection
coefficients $L(\lambda,t)$ and $R(\lambda,t)$
appearing in (4.4) and (4.6) are both identically zero.}
\item{(d)} {\it The transmission coefficients can be written as the ratio
of two determinants as}
$$T_{\text l}(\lambda)=\displaystyle\frac{\det(\lambda I+iA^\dagger)}
{\det(\lambda I-iA)},\quad
T_{\text r}(\lambda)=\displaystyle\frac{\det(\lambda I+iA^\dagger)}
{\det(\lambda I-iA)}.\tag 4.38$$

\noindent PROOF: Let us compare (4.4) and (4.5)
with (4.15) and (4.16) by ignoring
the subscript $\text{l}$ in
(4.15) and (4.16).
Using Proposition 4.1 (a) in the first
component of (4.15), we see that
$LT^{-1}_{\text l}=0.$ Then, with the help of (4.8) we also
conclude $RT^{-1}_{\text r}=0.$
Using Proposition 4.1 (c) in the second component of
(4.15) we obtain (4.34), and similarly by
using Proposition 4.1 (b) in the first component in (4.16) we obtain
$$(T_{\text r}^{-1})^\dagger=1+iB^\dagger N^{-1}(\lambda I-iA)^{-1}B.\tag 4.39$$
Applying Proposition 4.2 on (4.34) and (4.36) and simplifying the
resulting expressions with the help of (2.1), we obtain
(4.35) and (4.37). We then get (c) from
$LT^{-1}_{\text l}=0$ and $RT^{-1}_{\text r}=0.$ Applying Proposition 4.3
onto (4.35) and (4.37) and using (2.1)
we obtain (4.38). \qed

\vskip 10 pt
\noindent {\bf 5. THE MATRIX NLS EQUATION}
\vskip 3 pt

In this section we show that
all the results presented in Sections 2-4 for the scalar NLS equation
remain valid for the matrix NLS equation as well. In other words,
when the scalar function $u(x,t)$ in (1.3) is replaced with
an $m\times n$ matrix-valued function of $x$ and $t,$
our results in the previous sections remain valid.

In Section 2 we have started with the triplet $(A,B,C),$
where the sizes of $A,$ $B,$ $C$ were
$p\times p,$ $p\times 1,$ and $1\times p,$ respectively.
In Sections 2-4 we have carefully stated all our
results so that they all will remain valid also for the $m\times n$ matrix
NLS equation in (1.3).
In order for the reader to see the validity of all the results
in Sections 2-4 in the matrix case, in this section
we will use subscripts to indicate matrix dimensions by writing
$A$ as $A_{pp},$ $B$ as $B_{pm},$ $C$ as $C_{np},$ etc.
At appropriate places by interpreting $1$ as the identity matrix
$I_m$ or $I_n,$ and interpreting
$0$ as one of the four zero matrices $0_{mm},$ $0_{nn},$ $0_{mn},$
$0_{nm},$ we will show that
all the results in Sections 2-4 remain valid also in the matrix case.

To produce exact solutions to the $m\times n$ matrix NLS equation (1.3),
we start with a triplet
$(A_{pp},B_{pm},C_{np}).$
The Lyapunov equations in (2.1) remain unchanged
with $Q_{pp}$ and $N_{pp}$ as the solutions.
Theorem 2.1 and its proof remain unchanged, with the matrices
$F_{pp}$ and $u_{mn}$ still defined as in (2.2) and (2.3), respectively.
Note that $v(x,t)$ defined in (2.5) now becomes an $n\times m$ matrix, and
Theorem 2.1 remains valid with $v_{nm}$ as in (2.5) and $G_{pp}$ as given in (2.4)
for the $n\times m$ matrix NLS equation
$$i\displaystyle\frac{\partial v}{\partial t}+\displaystyle\frac{\partial^2 v}
{\partial x^2}+2vv^\dagger v=0.$$
Theorem 2.2 and its proof hold verbatim with $Q_{pp},$ $N_{pp},$ $F_{pp},$ $G_{pp}$
in the matrix case. Alternatively,
in defining $v$ one can simply use a triplet $(A,B,C)$ with sizes
$p\times p,$ $p\times n,$ $m\times p,$ respectively, so that
$v$ has the same size as $u,$ in which case Theorem 2.1 and its proof hold verbatim
for the $m\times n$ matrix NLS equation. In fact, this alternative is preferred, because in that
case
the generalizations of all the results in Sections 2-4 to the
matrix case are simpler to state.
Theorem 2.3 has a counterpart in the matrix case,
but we will not consider that generalization in this paper.

In terms of $u_{mn},$ $v_{nm},$ $\tilde u_{mn},$ and $\tilde v_{nm},$
all the results in Section 3 until Theorem 3.3 also
hold verbatim in the matrix case. Theorem 3.3
and its proof also hold
when stated with $u_{mn},$ $v_{nm},$ $\tilde u_{nm},$ and $\tilde v_{mn}.$
Alternatively, in evaluating
$v$ and $\tilde v,$ by using a triplet $(A,B,C)$
with sizes
$p\times p,$ $p\times n,$ $m\times p,$ respectively, all the results
in Section 3, including
Theorem 3.3 and its proof, hold verbatim in the matrix case.

All the results in Section 4 also hold with obvious interpretations
of the stated quantities. The matrix Zakharov-Shabat system corresponding
to (1.2) now becomes (1.4).
The Jost solutions
$\psi_{(m+n)n},$ $\phi_{(m+n)m},$ $\bar\psi_{(m+n)m},$ $\bar\phi_{(m+n)n}$ are now
matrices. In analogy with (4.2) and (4.3), their
asymptotics in the matrix case are given by
$$\psi(\lambda,x,t)=\bmatrix 0_{mn}\\
\noalign{\medskip}
e^{i\lambda x}I_n\endbmatrix [1+o(1)],
\quad
\bar\psi(\lambda,x,t)=\bmatrix e^{-i\lambda x}I_m\\
\noalign{\medskip}
0_{nm}\endbmatrix [1+o(1)],
\qquad x\to+\infty,$$
$$\phi(\lambda,x,t)=\bmatrix e^{-i\lambda x}I_m\\
\noalign{\medskip}
0_{nm}\endbmatrix [1+o(1)],
\quad
\bar\phi(\lambda,x,t)=\bmatrix 0_{mn}\\
\noalign{\medskip}
e^{i\lambda x}I_n\endbmatrix [1+o(1)],
\qquad x\to-\infty.$$
The definition of the Wronskian in (4.1) holds in the matrix case, where it is now
a square matrix of size either $m\times m$ or $n\times n,$ depending on
the sizes of matrix solutions.
The asymptotic
relations (4.4)-(4.7) defining
the scattering coefficients hold as stated, where now the scattering coefficients
$R_{nm},$ $L_{mn},$ $(T_{\text l})_{nn},$ and $(T_{\text r})_{mm}$ are matrices. The Fourier
relations (4.9) and (4.10) are also valid in the matrix case and are given by
$$\psi(\lambda,x,t)=\bmatrix
0_{mn}\\
\noalign{\medskip}
e^{i\lambda x}I_n\endbmatrix+\int_x^\infty dy\,K(x,y,t)\,e^{i\lambda y},$$
$$\bar\psi(\lambda,x,t)=\bmatrix
e^{-i\lambda x}I_m\\
\noalign{\medskip}
0_{nm}
\endbmatrix+\int_x^\infty dy\,\bar K(x,y,t)\,e^{-i\lambda y},$$
where $K(x,y,t)$ is now an $(m+n)\times n$ matrix
and $\bar K(x,y,t)$
is an $(m+n)\times m$ matrix. The Marchenko equations
in (4.12) and (4.13) remain unchanged with the understanding
that the quantity $\Omega_{\text l}(y,t)$ appearing in the kernel
is now an $n\times m$ matrix and we have
$$\bar K(x,y,t)+\bmatrix 0_{mm}\\
\noalign{\medskip}
\Omega_{\text l}(x+y,t)\endbmatrix+\int_x^\infty dy\,K(x,z,t)\,\Omega_{\text l}(z+y,t)=\bmatrix 0_{mm}\\
\noalign{\medskip}
0_{nm}\endbmatrix,\qquad y>x,$$
$$K(x,y,t)-\bmatrix \Omega_{\text l}(x+y,t)^\dagger\\
\noalign{\medskip}
0_{nn}\endbmatrix-\int_x^\infty dy\,\bar K(x,z,t)\,\Omega_{\text l}(z+y,t)^\dagger=\bmatrix 0_{mn}\\
\noalign{\medskip}
0_{nn}\endbmatrix,\qquad  y>x,$$
and the potential $u_{mn}$ is recovered as in (4.11) via
$$u(x,t)=-2\bmatrix I_m&0_{mn}\endbmatrix K(x,x,t)=2\,\bar K(x,x,t)^\dagger \bmatrix 0_{mn}\\
\noalign{\medskip}
I_n\endbmatrix.$$

Similarly,
the Fourier representations in (4.17) and (4.18)
remain true by interpreting them as
$$\phi(\lambda,x,t)=\bmatrix
e^{-i\lambda x}I_m\\
\noalign{\medskip}
0_{nm}
\endbmatrix+\int_{-\infty}^x dy\,M(x,y,t)\,e^{-i\lambda y},$$
$$\bar\phi(\lambda,x,t)=\bmatrix
0_{mn}\\
\noalign{\medskip}
e^{i\lambda x}I_n
\endbmatrix+\int_{-\infty}^x dy\,\bar M(x,y,t)\,e^{i\lambda y},$$
where now $M(x,y,t)$ and $\bar M(x,y,t)$ have
sizes $(m+n)\times m$ and $(m+n)\times n,$ respectively.

In generalizing (4.22) to the matrix case, let us
simply use a triplet
$(A_{\text r},B_{\text r},C_{\text r})$ with sizes
$p\times p,$ $p\times n,$ $m\times p,$ respectively.
With the understanding that $\Omega_{\text r}(y,t)$ is
now an $m\times n$ matrix,
the Marchenko equations (4.20) and (4.21)
hold true and can also be written as
$$\bar M(x,y,t)+\bmatrix \Omega_{\text r}(x+y,t)\\
\noalign{\medskip}
0_{nn}\endbmatrix+\int_{-\infty}^x dy\,M(x,z,t)\,\Omega_{\text r}(z+y,t)=\bmatrix 0_{mn}\\
\noalign{\medskip}
0_{nn}\endbmatrix,\qquad y<x,$$
$$M(x,y,t)-\bmatrix 0_{mm}\\
\noalign{\medskip}
\Omega_{\text r}(x+y,t)^\dagger\endbmatrix
-\int_{-\infty}^x dy\,\bar M(x,z,t)\,\Omega_{\text r}(z+y,t)^\dagger=\bmatrix 0_{mm}\\
\noalign{\medskip}
0_{nm}\endbmatrix,\qquad  y<x,$$
and the potential $v_{mn}$ is recovered as in (4.19) via
$$v(x,t)=2\bmatrix I_m&0_{mn}\endbmatrix \bar M(x,x,t)=-2\,M(x,x,t)^\dagger \bmatrix 0_{mn}\\
\noalign{\medskip}
I_n\endbmatrix.$$

When the triplet $(A_{\text l},B_{\text l},C_{\text l})$
with matrix sizes $p\times p,$ $p\times m,$ $n\times p,$ respectively,
is used in
the Marchenko kernel (4.14), the
Jost solutions $\psi_{(m+n)n}$ and $\bar \psi_{(m+n)m}$ on any half line $x\in(a,+\infty)$
for any real number $a$ can be expressed as in
(4.15) and (4.16) via
$$\psi(\lambda,x,t)=\bmatrix iB_{\text l}^\dagger
F_{\text l}(x,t)^{-1}(\lambda I+iA_{\text l}^\dagger)^{-1}
e^{i\lambda x}
C_{\text l}^\dagger\\
\noalign{\medskip}
e^{i\lambda x}I_n-iC_{\text l}[F_{\text l}(x,t)^\dagger]^{-1}N_{\text l}
e^{-2A_{\text l}^\dagger x+4i(A_{\text l}^\dagger)^2t}
(\lambda I+iA_{\text l}^\dagger)^{-1}e^{i\lambda x}
C_{\text l}^\dagger
\endbmatrix,$$
$$\bar\psi(\lambda,x,t)=\bmatrix
e^{-i\lambda x}I_m+iB_{\text l}^\dagger
e^{-2A_{\text l}^\dagger x+4i(A_{\text l}^\dagger)^2t}
Q_{\text l}
[F_{\text l}(x,t)^\dagger]^{-1}
(\lambda I-iA_{\text l})^{-1}e^{-i\lambda x}B_{\text l}\\
\noalign{\medskip}
iC_{\text l}
[F_{\text l}(x,t)^\dagger]^{-1}
(\lambda I-iA_{\text l})^{-1}e^{-i\lambda x}B_{\text l}
\endbmatrix.$$

Similarly, when the matrix triplet $(A_{\text r},B_{\text r},C_{\text r})$
of sizes $p\times p,$ $p\times n,$ $m\times p,$ respectively,
is used in
the Marchenko kernel (4.22), the corresponding
Jost solutions $\phi_{(m+n)m}$ and $\bar \phi_{(m+n)n}$ on any half line $x\in(-\infty,a)$
for any real number $a$ can be expressed as in
(4.23) and (4.24) via
$$\phi(\lambda,x,t)=\bmatrix e^{-i\lambda x}I_m-i
C_{\text r}
G_{\text r}(x,t)^{-1}
N_{\text r}
e^{2A_{\text r}^\dagger x-4i(A_{\text r}^\dagger)^2t}
(\lambda I+iA_{\text r}^\dagger)^{-1}e^{-i\lambda x}C_{\text r}^\dagger\\
\noalign{\medskip}
iB_{\text r}^\dagger[G_{\text r}(x,t)^\dagger]^{-1}
(\lambda I+iA_{\text r}^\dagger)^{-1}e^{-i\lambda x}C_{\text r}^\dagger\endbmatrix,
$$
$$\bar\phi(\lambda,x,t)=\bmatrix
iC_{\text r}
G_{\text r}(x,t)^{-1}
(\lambda I-iA_{\text r})^{-1}
e^{i\lambda x}B_{\text r}\\
\noalign{\medskip}
e^{i\lambda x}I_n+
iB_{\text r}^\dagger
[G_{\text r}(x,t)^\dagger]^{-1}
Q_{\text r}
e^{2A_{\text r} x+4iA_{\text r}^2t}
(\lambda I-iA_{\text r})^{-1}
e^{i\lambda x}
B_{\text r}
\endbmatrix.$$

Proposition 4.1 and its proof hold as stated with the understanding that
we interpret $0$ in (a) as $0_{pp}.$ Proposition 4.2
already contains the matrix case in (4.30), and
Proposition 4.3 is already stated in such a way that it can directly
be used in the matrix case.
Theorem 4.4 and its proof in the matrix case hold with the following minor
differences. The quantities $u,$ $T_{\text l},$ $T_{\text r},$
$L,$ $R$ are now matrices
$u_{mn},$ $(T_{\text l})_{nn},$ $(T_{\text r})_{mm},$
$L_{mn},$ $R_{nm},$ respectively;
(4.34)-(4.37) and (4.39) hold as
stated with the understanding that the scalar $1$ becomes an
identity matrix $I$ of appropriate size, namely,
$$T_{\text l}^{-1}=I_n-iCQ^{-1}
(\lambda I_p+iA^\dagger)^{-1}C^\dagger,$$
$$T_{\text l}=I_n+iC(\lambda I_p-iA)^{-1}Q^{-1}C^\dagger,$$
$$T_{\text r}^{-1}=I_m-iB^\dagger
(\lambda I_p+iA^\dagger)^{-1} N^{-1}B,$$
$$T_{\text r}=I_m+iB^\dagger N^{-1}(\lambda I_p-iA)^{-1}B.$$
In the matrix case, on the left hand sides of the two equations in (4.38) we need to replace
$T_{\text l}$ and $T_{\text r}$ by their determinants; namely,
(4.38) in the matrix case becomes
$$\det T_{\text l}(\lambda)=\displaystyle\frac{\det(\lambda I_p+iA^\dagger)}
{\det(\lambda I_p-iA)},\quad
\det T_{\text r}(\lambda)=\displaystyle\frac{\det(\lambda I_p+iA^\dagger)}
{\det(\lambda I_p-iA)}.$$

\vskip 10 pt
\noindent {\bf 6. AN EXAMPLE}
\vskip 3 pt

We conclude our paper by providing an application of Theorem 3.2 to a specific case.
In Example 7.2 of [3] we evaluated the exact solution to the NLS equation
corresponding to the triplet $(A,B,C)$ with
$$A=\bmatrix 2&0\\
\noalign{\medskip}
0&-1\endbmatrix,\quad B=\bmatrix 1\\
\noalign{\medskip}
1\endbmatrix,\quad C=\bmatrix 1&-1\endbmatrix.\tag 6.1$$
We incorrectly conjectured in that example that we had a nonsoliton solution
because one of the eigenvalues, of $A$ was not positive. Using the result of Theorem 3.2 of
the present paper, we are now able to confirm that that solution is indeed
a two-soliton solution. For this, we proceed as follows.

Using the triplet $(A,B,C)$ of (6.1), we solve (2.1)
in a straightforward manner and get
$$Q=\bmatrix 1/4&-1\\
\noalign{\medskip}
-1&-1/2\endbmatrix,\quad N=\bmatrix 1/4&1\\
\noalign{\medskip}
1&-1/2\endbmatrix.$$
We then construct $(\tilde A,\tilde B,\tilde C,\tilde Q,\tilde N,\tilde F)$
via (3.9)-(3.13) and obtain
$$\tilde A=
\bmatrix 2&0\\
\noalign{\medskip}
0&1\endbmatrix,\quad \tilde B=\bmatrix 3\\
\noalign{\medskip}
2\endbmatrix,\quad \tilde C=\bmatrix 3&-2\endbmatrix,\tag 6.2$$
$$\tilde Q=\bmatrix 9/4&-2\\
\noalign{\medskip}
-2&2\endbmatrix,\quad \tilde N=\bmatrix 9/4&2\\
\noalign{\medskip}
2&2\endbmatrix,$$
$$\tilde F=\bmatrix e^{4x-16it}+\displaystyle\frac{81}{16}\,e^{-4x-16it}-4e^{-2x-4it}&
\displaystyle\frac{9}{2}\,e^{-4x-16it}-4e^{-2x-4it}\\
\noalign{\medskip}
-\displaystyle\frac{9}{2}\,e^{-4x-16it}+4e^{-2x-4it}&e^{2x-4it}-4e^{-4x-16it}+4e^{-2x-4it}
\endbmatrix.$$
As seen from (6.2), the eigenvalue $(-1)$ of $A$ is transformed into the eigenvalue
$(+1)$ of $\tilde A.$
The potential $\tilde u(x,t),$ or equivalently $u(x,t),$ is then constructed
via (2.3) and we get
$$u(x,t)=\displaystyle\frac{8e^{4it}(9e^{-4x}+16e^{4x})-32e^{16it}(4e^{-2x}+9e^{2x})}{
-128\cos(12 t)+4e^{-6x}+16e^{6x}+81e^{-2x}+64e^{2x}},$$
agreeing with the potential of Example 7.2 of [3].
For this potential, the transmission coefficients are evaluated via (4.38) as
$$T_{\text l}(\lambda)=T_{\text r}(\lambda)=\displaystyle\frac{(\lambda+2i)(\lambda+i)}{
(\lambda-2i)(\lambda-i)},$$
because the real parts of all eigenvalues of
$\tilde A$ in the associated triplet $(\tilde A,\tilde B,\tilde C)$ are positive.
As for the norming constants associated with the bound states at $\lambda=2i$
and $\lambda=i,$ we need to transform the triplet $(\tilde A,\tilde B,\tilde C)$ of (6.2)
into another triplet $(A,B,C),$ different from (6.1), so that
we will have $B=\bmatrix 1\\
1\endbmatrix.$ By using (2.16) we obtain
$$M=\bmatrix 1&0\\
\noalign{\medskip}
0&1\endbmatrix,\quad S=\bmatrix 3&0\\
\noalign{\medskip}
0&2\endbmatrix,\quad A=
\bmatrix 2&0\\
\noalign{\medskip}
0&1\endbmatrix,\quad B=\bmatrix 1\\
\noalign{\medskip}
1\endbmatrix,\quad C=\bmatrix 9&-4\endbmatrix.$$
Thus, the norming constant at $\lambda=2i$ is $9$ and the norming constant at
$\lambda=i$ is $-4.$
Finally, the Jost solutions $\psi(\lambda,x,t)$ and $\bar\psi(\lambda,x,t)$ to the
Zakharov-Shabat system (1.2) are obtained via (4.15) and (4.16),
respectively, by using
$(\tilde A,\tilde B,\tilde C,\tilde Q,\tilde N,\tilde F)$ for
$(\tilde A_{\text l},\tilde B_{\text l},\tilde C_{\text l},
\tilde Q_{\text l},\tilde N_{\text l},\tilde F_{\text l})$
there. For example, for the Jost solution $\psi$ we get
$$\psi(\lambda,x,t)=\bmatrix 0\\
\noalign{\medskip} e^{i\lambda x}\endbmatrix
+\displaystyle\frac{\bmatrix 4 i e^{-4 x+4it}g_1(\lambda,x,t)\\
\noalign{\medskip} 4i\,g_2(\lambda,x,t)\endbmatrix
e^{i\lambda x}}{(\lambda+2i)(\lambda+i)[-128\cos(12 t)+4e^{-6x}+16e^{6x}+81e^{-2x}+64e^{2x}]},$$
where we have defined
$$g_1(\lambda,x,t):=
36(\lambda+i)e^{6x+12it}+16(\lambda-i)e^{2x+12it}-16
(\lambda+2i)e^{8x}-9(\lambda-2i),$$
$$g_2(\lambda,x,t):=48(\lambda+i)e^{12it}+
48(\lambda+2i)e^{-12it}-6\lambda e^{-6x}-81
(\lambda+i)e^{-2x}-32(\lambda+2i)e^{2x}.$$

\vskip 5 pt

\noindent{\bf Acknowledgment}.
One of the authors (T.A.) is greatly indebted to the University of Cagliari for its
hospitality during a recent visit.
The research leading to this article was supported in part
by the National Science Foundation under grant DMS-0610494,
INdAM, and the Autonomous Region of Sardinia.

\vskip 5 pt

\noindent {\bf REFERENCES}

\item{[1]} M. J. Ablowitz and P. A. Clarkson, {\it Solitons, nonlinear
evolution equations and inverse scattering,} Cambridge Univ. Press, Cambridge,
1991.

\item{[2]} M. J. Ablowitz and H. Segur, {\it
Solitons and the inverse scattering
transform,} SIAM, Philadelphia, 1981.

\item{[3]} T. Aktosun, F. Demontis, and C. van der Mee,
{\it Exact solutions to the focusing nonlinear Schr\"o\-din\-ger equation},
Inverse Problems {\bf 23}, 2171--2195 (2007).

\item{[4]} T. Aktosun and C. van der Mee, {\it Explicit solutions to the
Korteweg-de Vries equation on the half-line,} Inverse Problems {\bf 22},
2165--2174 (2006).

\item{[5]} H. Bart, I. Gohberg, and M. A. Kaashoek, {\it Minimal
factorization of matrix and operator functions,} Birkh\"auser,
Basel, 1979.

\item{[6]} F. Demontis, {\it Direct and inverse scattering of the matrix
Zakharov-Shabat system}, Ph.D. thesis, University of Cagliari, Italy, 2007.

\item{[7]} F. Demontis and C. van der Mee, {\it Marchenko equations
and norming constants of the matrix Zakharov-Shabat system}, Operators and
Matrices {\bf 2}, 79--113 (2008).

\item{[8]} F. Demontis and C. van der Mee, {\it Explicit solutions
of the cubic matrix nonlinear Schr\"o\-din\-ger equation}, Inverse Problems
{\bf 24}, 02520 (2008), 16 pp.

\item{[9]} H. Dym, {\it Linear algebra in action}, Graduate Studies in Mathematics,
78, Amer. Math. Soc., Providence, RI, 2007.

\item{[10]} A. Hasegawa and M. Matsumoto, {\it Optical solitons in fibers,}
3rd ed., Springer, Berlin, 2002.

\item{[11]} A. Hasegawa and F. Tappert, {\it Transmission of stationary
nonlinear optical pulses in dispersive dielectric fibers. I. Anomalous
dispersion,} Appl. Phys. Lett. {\bf 23}, 142--144 (1973).

\item{[12]} A. Hasegawa and F. Tappert, {\it Transmission of stationary
nonlinear optical pulses in dispersive dielectric fibers. II. Normal
dispersion,} Appl. Phys. Lett. {\bf 23}, 171--172 (1973).

\item{[13]} S. Novikov, S. V. Manakov, L. P. Pitaevskii, and V. E. Zakharov,
{\it Theory of solitons,}
Consultants Bureau, New York, 1984.

\item{[14]} V. E. Zakharov, {\it Stability of periodic waves of
finite amplitude on the surface of a deep fluid,}
J. Appl. Mech. Tech. Phys. {\bf 4}, 190--194 (1968).

\item{[15]} V. E. Zakharov and A. B. Shabat, {\it Exact theory of
two-dimensional self-focusing and one dimensional self-modulation of waves in
nonlinear media,} Sov. Phys. JETP {\bf 34}, 62--69 (1972).

\end